\documentclass[conference]{IEEEtran}
\pdfoutput=1
\usepackage{graphicx}
\usepackage{epstopdf}
\ifCLASSOPTIONcompsoc
  \usepackage[caption=false,font=normalsize,labelfont=sf,textfont=sf]{subfig}
\else
  \usepackage[caption=false,font=footnotesize]{subfig}
\fi
\usepackage{booktabs} 

\usepackage{hyphenat}
\usepackage{microtype}
\usepackage{url}

\usepackage{amsmath,amssymb,amsfonts}
\usepackage{cleveref}
\crefformat{section}{\S#2#1#3}
\crefformat{subsection}{\S#2#1#3}
\crefformat{subsubsection}{\S#2#1#3}

\usepackage{units}
\usepackage{boldline}
\usepackage{tabularx}
\newcolumntype{R}{>{\raggedleft\arraybackslash}X}
\usepackage{colortbl}

\usepackage[table,xcdraw]{xcolor}
\usepackage{multirow}
\usepackage{comment}

\usepackage{enumitem}
\setlist{leftmargin=5.5mm}

\usepackage[textsize=tiny]{todonotes}
\usepackage[skip=0pt,font=small]{caption}

\providecommand{\DIFdel}[1]{} 

\newcommand{\memo}[1]{\textcolor{purple}{#1}}

\newcommand{\cC}[0]{\cellcolor[HTML]{CCCCCC}}
\newcommand{\rC}[0]{\rowcolor[HTML]{CCCCCC}}
\newcommand{\hC}[0]{\rowcolor[HTML]{333333}}
\newcommand{\tH}[1]{\multicolumn{1}{c}{\textcolor{white}{#1}}}

\newcommand{\TSS}[1]{\textsuperscript{#1}}

\usepackage{etoolbox}

\newtoggle{inclIEEECopyRight}
\toggletrue{inclIEEECopyRight}

\begin{document}
\iftoggle{inclIEEECopyRight}{
    \begin{titlepage}
    \mbox{}\\{\Large \textbf{IEEE Copyright Notice}}
    \newline\newline\newline\newline
    \textcopyright~2021 IEEE. Personal use of this material is permitted.
    Permission from IEEE must be obtained for all other uses, in any current
    or future media, including reprinting/republishing this material for
    advertising or promotional purposes, creating new collective works, for
    resale or redistribution to servers or lists, or reuse of any copyrighted
    component of this work in other works.
    \newline\newline\newline\newline
    {\large Accepted to be Published in: Proceedings of the 35rd IEEE
    International Parallel \& Distributed Processing Symposium, May 17-21,
    2021 Portland, Oregon, USA}
    \end{titlepage}
}{}

\bstctlcite{IEEEexample:BSTcontrol}

\title{Matrix Engines for High Performance Computing: \\\huge A Paragon of Performance or Grasping at Straws?\vspace{-1.0em}}  


\author{
    \IEEEauthorblockN{Jens Domke\IEEEauthorrefmark{1}\IEEEauthorrefmark{4},
	Emil Vatai\IEEEauthorrefmark{1}\IEEEauthorrefmark{4},
	Aleksandr Drozd\IEEEauthorrefmark{1}\IEEEauthorrefmark{4},
	Peng Chen\IEEEauthorrefmark{2},
	Yosuke Oyama\IEEEauthorrefmark{4},
	Lingqi Zhang\IEEEauthorrefmark{4},\\
	Shweta Salaria\IEEEauthorrefmark{1}\IEEEauthorrefmark{4},
	Daichi Mukunoki\IEEEauthorrefmark{1},
	Artur Podobas\IEEEauthorrefmark{3},
	Mohamed Wahib\IEEEauthorrefmark{2}\IEEEauthorrefmark{1},
	Satoshi Matsuoka\IEEEauthorrefmark{1}\IEEEauthorrefmark{4}
}

\IEEEauthorblockA{
    \small{
      \IEEEauthorrefmark{1}
      RIKEN CCS, Japan
      \texttt{\{jens.domke,emil.vatai,aleksandr.drozd,shweta.salaria,daichi.mukunoki\}@riken.jp}
    }
  }
	  \IEEEauthorblockA{
    \small{\parbox{\linewidth}{\centering
      \IEEEauthorrefmark{2}
      National Institute of Advanced Industrial Science and Technology, Japan
      \texttt{\{mohamed.attia,chin.hou\}@aist.go.jp}
    }}}
  \IEEEauthorblockA{
    \small{
      \IEEEauthorrefmark{3}
      KTH Royal Institute of Technology, Stockholm, Sweden
      \texttt{podobas@kth.se}
    }
  }
  \IEEEauthorblockA{
    \small{
      \IEEEauthorrefmark{4}
      Tokyo Institute of Technology, Japan
      \texttt{\{oyama.y.aa,zhang.l.ai\}@m.titech.ac.jp}\vspace{-.5em}
    }
  }
}

\maketitle
\thispagestyle{plain}
\pagestyle{plain}

\begin{abstract}

Matrix engines or units, in different forms and affinities, are becoming a
reality in modern processors; CPUs and otherwise. The current and dominant
algorithmic approach to Deep Learning merits the commercial investments in
these units, and deduced from the No.~1 benchmark in supercomputing, namely
High Performance Linpack, one would expect an awakened enthusiasm by the
HPC community, too.

Hence, our goal is to identify the practical added benefits
for HPC and machine learning applications by having access to matrix engines.
For this purpose, we perform an in-depth survey of software stacks, proxy
applications and benchmarks, and historical batch job records. We provide
a cost-benefit analysis of matrix engines, both asymptotically and in
conjunction with state-of-the-art processors. While our empirical data will
temper the enthusiasm, we also outline opportunities to ``misuse'' these
dense matrix-multiplication engines if they come for free.
\end{abstract}

\IEEEpeerreviewmaketitle

\section{Introduction}\label{sec:intro} 
With both Dennard's scaling~\cite{dennard1974design} and Moore's law~\cite{schaller1997moore} gone, computer scientists and architects are perhaps facing their grandest challenge to date. Today, computer scientists are actively chasing Post-Moore alternatives such as the intrusive neuromorphic and quantum computers~\cite{vetter2017architectures}. However, not all options need to be intrusive, and some merely require us to move away from traditional von-Neumann architectures. Among the more salient of these options is architectural specialization~\cite{shalf2020future}. Hardware specialization focuses on accelerating application-specific core components to reduce the needless energy tax~\cite{hameed2010understanding} that a traditional von-Neumann general-purpose system demands. Instead, the aspiration is to maximize data locality and fully eliminate the operation control cost that is continuously present in GPUs (e.g., instruction fetching and decoding, etc.)~\cite{jouppi2018motivation}. Architecture-specialization is not a new concept in itself, where co-processors and accelerators based on Field-Programmable Gate Arrays (FPGAs)~\cite{kuon2008fpga,zohouri2016evaluating}, Coarse-Grained Reconfigurable Architectures (CGRAs)~\cite{podobas2020survey}, or Application-Specific Integrated Circuits (ASICs) (e.g., Anton~\cite{shaw2008anton} or Grape~\cite{makino2012grape}) have continuously accompanied computer systems in their historical road to performance. 

Among the more popular candidates for architecture specialization, much thanks to the limitless popularity of Deep-Learning~\cite{lecun2015deep}, is to target General Matrix Multiplication (GEMM). Targeting GEMM is perhaps not entirely unmotivated: GEMM is often \textit{claimed} to be the core compute-intensive component in many scientific applications spanning multiple domains, such as Computational Fluid Dynamics (e.g., NEK5000~\cite{shin2010speeding}) or Deep Learning (DL)~\cite{warden2015gemm}. Today, there are already a large bulk of \textit{application-specific} (primarily~DL) accelerators that are based around systolic arrays~\cite{kung1978systolic} (essentially GEMM engines), such as Huawei's Ascend 910~\cite{liao2019davinci} and Google Tensor Processing Units (TPUs)~\cite{jouppi2017datacenter}.

More importantly, the trend of adopting hardware acceleration for GEMM operations is coming even to \textit{general-purpose architectures} and their Instruction Set Architecture (ISA). NVIDIA introduced the Tensor Cores~\cite{choquette2018volta} in the Volta, Ampere, and Turing series of accelerators. Both Intel (with Sapphire Rapids~\cite{intelamx}) and IBM (with Power10~\cite{starke_ibms_2020}) are extending their SIMD-capabilities to support matrix operations, with similar proposals by authors dating back a decade~\cite{soliman2007mat}. The unspoken question is: \textit{Is the inclusion of specialized matrix engines in general-purpose processors truly motivated and merited, or is the silicon better invested in other parts?}

In this paper, we aspire to holistically look at the inclusion of matrix engines---abbreviated \textbf{ME} hereafter---into the general-purpose processor and its expected impact on High-Performance Computing (HPC) applications. It is important to emphasize that we consider DL as one of many workloads in HPC, and not the application we solely focus on. In this study, we target to answer the following three questions:
\begin{itemize}[leftmargin=2.5mm]
	\item Does the occurrence and usage of matrix operations in scientific workloads truly merit matrix engines' inclusion into general-purpose ISAs?
	\item What performance benefits can we expect from using MEs on existing scientific applications that can leverage them?
	\item Performance projection of using matrix engines on future scientific workloads using a model empirically derived from the NVIDIA V100 GPUs.
\end{itemize}
To answer the above questions, our contributions are:
\begin{itemize}[leftmargin=2.5mm]
	\item We inspect software management packages, historical batch job records, profiles, and source code of a board set of HPC and Machine Learning proxy applications and benchmarks to identify dense matrix requirements.
	\item We provide a cost-benefit analysis of projected performance gains from matrix engines, driven by resource usage per domain in different production supercomputers.
	\item A detailed discussion of opportunities and challenges in adopting matrix engines, from the perspective of HPC workloads.
\end{itemize}

\section{Matrix Engines from a Hardware Perspective}\label{sec:hardware} 

\begin{table*}[tbp]
  \caption{Overview of existing and emerging general-purpose and AI architectures that leverage matrix engines to accelerate computations (\textbf{f16} = IEEE-754 16-bit or BFloat16, \textbf{f32} = IEEE-754 single prec., \textbf{f64} = IEEE-754 double prec.; \textbf{GF} = \unit{Gflop/s}; INT4/8 support omitted\big)}
    \centering\scriptsize
    \begin{tabularx}{\textwidth}{ |c|X|r|r|c|r|r|r|c| }
    \hline \hC
    \tH{\small Type} & \tH{\small System}  & \tH{\small Tech.} & \tH{\small Die size} & \tH{\small ME size} & \tH{\small \unit{Tflop/s} (f16)} & \tH{\small \unit{Tflop/s} (f32)} & \tH{\small \unit{Tflop/s} (f64)} & \tH{\small Support}  \\ \hline
    \multirow{4}{*}{\rotatebox[origin=c]{90}{General}}
     &  Intel Sapphire Rapids\footnotemark & \unit[10]{nm} & \multicolumn{1}{c|}{|} & 16x32 & \multicolumn{1}{c|}{|}  & \multicolumn{1}{c|}{|} & \multicolumn{1}{c|}{|} &  f16\\  \cline{2-9}
     &\cC  IBM Power10~\footnotemark        &\cC  \unit[7]{nm}   &\cC \unit[602]{mm\TSS{2}} &\cC  4x4     &\cC 16.4~~(\unit[27.2]{GF/mm\TSS{2}})  &\cC 8.2~(\unit[13.6]{GF/mm\TSS{2}})   &\cC 4.1~~(\unit[6.8]{GF/mm\TSS{2}}) &\cC f16, f32, f64 \\ \cline{2-9}
     &  NVIDIA Tesla V100     & \unit[12]{nm}   & \unit[815]{mm\TSS{2}} & 4x4x4    & 125.0~(\unit[153.4]{GF/mm\TSS{2}})  & 15.7~(\unit[19.3]{GF/mm\TSS{2}})  & 7.8~~(\unit[9.6]{GF/mm\TSS{2}}) & f16 \\ \cline{2-9}
     &\cC  NVIDIA Tesla A100\footnotemark     &\cC  \unit[7]{nm}   &\cC \unit[826]{mm\TSS{2}} &\cC 4x4x4    &\cC 312.0~(\unit[377.7]{GF/mm\TSS{2}})  &\cC 19.5~(\unit[23.6]{GF/mm\TSS{2}}) &\cC 19.5~(\unit[23.6]{GF/mm\TSS{2}}) &\cC f16, f32, f64 \\ \hline\hline
    \multirow{4}{*}{\rotatebox[origin=c]{90}{AI}}
     &  Google TPUv2          & \unit[20]{nm}   & \multicolumn{1}{c|}{|} & 128x128  & 45.0~~(|)   & \multicolumn{1}{c|}{|} & \multicolumn{1}{c|}{|} & f16 \\ \cline{2-9}
     &\cC  Google TPUv3          &\cC \unit[16]{nm}   & \multicolumn{1}{c|}{\cC |} &\cC 128x128  &\cC 90.0~~(|)    & \multicolumn{1}{c|}{\cC |} & \multicolumn{1}{c|}{\cC |} &\cC f16 \\ \cline{2-9}
     &  Habana Labs Gaudi     & \unit[16]{nm}   & \unit[500]{mm\TSS{2}} & Shared   & \multicolumn{1}{c|}{|} & \multicolumn{1}{c|}{|} & \multicolumn{1}{c|}{|} & f16, f32 \\ \cline{2-9}
     &\cC  Huawei Ascend 910\footnotemark     &\cC  \unit[7]{nm} &\cC \unit[1228]{mm\TSS{2}} &\cC 16x16x16 &\cC 256.0~(\unit[208.5]{GF/mm\TSS{2}})   & \multicolumn{1}{c|}{\cC |} & \multicolumn{1}{c|}{\cC |} &\cC f16 \\ \hline
    \end{tabularx}
    \label{tab:overview}
    \vspace{-4ex}
\end{table*} 
\addtocounter{footnote}{-3}\footnotetext{AMX listed for completeness (16x64 ME for INT8); performance unknown.}
\addtocounter{footnote}{+1}\footnotetext{Performance is calculated assuming 16 SMT8 cores running at \unit[4]{GHz}.}
\addtocounter{footnote}{+1}\footnotetext{The A100 also offers a unique hybrid 19-bit TF32 format combining a 10-bit mantissa of IEEE-754 half-precision and 8-bit exponent of BFloat16, yielding up to \unit[156]{Tflop/s} with TCs.}
\addtocounter{footnote}{+1}\footnotetext{Die size includes the Nimbus co-accelerator and four HBM2 stacks.}

In this section, we describe historical and current developments of ME adoption by vendors. The section is structured as follows: first, we discuss the historical lead up to MEs, then elaborate on current trends in hardware, and finally, we highlight the HPC community's motivation for leveraging MEs.

\subsection{Matrix Engines Lineage}\label{ssec:MElinage}
Early processors were scalar: executing a single operation on pairs of operands one at a time, possibly using Out-of-Order (OoO) and superscalar execution to improve performance via increased Instruction Level Parallelism (ILP)~\cite{wall1991limits}.
As time moved on, Moore's law facilitated the inclusion of more complex BLAS-1 (Vector-Vector) operations in hardware through the use of Single Instruction Multiple Data (SIMD)~\cite{duncan1990survey} units.
These additions, refined in systems such as ARM A64FX's Scalable Vector Extensions (SVE)~\cite{yoshida2018fujitsu}, Intel's AVX512~\cite{cornea2015intel}, or abstracted away through threading in NVIDIA's CUDA~\cite{kirk2007nvidia}, did increase the performance of processors at the cost of a small area of silicon; area that was available due to Moore's law.
Moving (or upgrading) these SIMD units to support GEMM operations is only the next natural step, with one small caveat: Moore's law is approaching its end~\cite{theis2017end}. The obvious question becomes: \textit{Are MEs really what we should be spending our silicon on, given that Moore's law is about to die out?}

\subsection{Matrix Engines: State-of-the-Art Performance and Trends}
We compile Table~\ref{tab:overview} to highlight some well known commercial general-purpose systems or AI accelerators that already exist or are about to be released (for the plethora of academic AI accelerators, we refer interested readers to surveys on the subject~\cite{reuther2019survey,podobas2020survey,wang2018survey}). All these new systems contain some form of ME integrated into the architecture.  From the general-purpose side, both the upcoming Intel Sapphire Rapids and the IBM Power10 will be augmented with MEs. According to public documents, IBM Power10~\cite{starke_ibms_2020} will be the more general CPU of the two, with support for a large variety of numerical representations (Int[4$|$8$|$16], FP[16$|$32$|$64]), while the Intel's AMX unit will focus more on AI applications (bfloat16), based on the available information in the programming reference~\cite{intelamx}. IBM Power10 features a hybrid model, which means that it accumulates into a wider representation than what it multiplies in; the only exception is for FP64, where it both multiplies and accumulates into the same representation.

From the GPU side, both NVIDIA~V100~\cite{choquette2018volta} and A100 feature MEs, called Tensor Cores, abbreviated TCs hereafter. The MEs in the V100 are hybrid, meaning that they accumulate into a wider numerical representation (in V100's case: FP32) than what it multiplies with (FP16). This limitation was overcome in A100, which supports up to double-precision (FP64) in its MEs.
From a performance perspective, the GPU-based systems---in particular the new A100---are grossly outperforming the other systems in both a higher peak performance (\unit{Tflop/s}) and a higher compute density (\unitfrac{Gflop/s~}{~mm\textsuperscript{2}}), where the IBM Power10 only reaches~18\% of the compute-density of an NVIDIA~V100. The NVIDIA~A100 is reported to reach up to \unit[312]{Tflop/s} of half-precision (FP16) performance.

The available information for AI-accelerators is more sparse and most architectures primarily report peak performance. Furthermore, most focus exclusively on bfloat16 (same as Intel Sapphire Rapids), and overall show a higher performance over general-purpose CPUs. The Habana Labs Gaudi~\cite{habana_labs_ltd_whitepaper_2019} architecture uses a shared ME unit accessible to all the cores, but most of the architectural details (including performance) are undisclosed. The highest performance and compute density of the AI accelerators delivers the Huawei's Ascend 910~\cite{liao2019davinci}, which reaches \unit[256]{Tflop/s} of raw FP16 performance, leading to \unitfrac[208]{Gflop/s~}{~mm\textsuperscript{2}} of compute density---nearly an order of magnitude~(7.7x) more than IBM~Power10 and but still only~55\% of the NVIDIA~A100's peak performance.

To summarize: modern architectures are moving towards integration of MEs into the core fabric.
Specializing the silicon towards acceleration of GEMM can be worthwhile, and can yield substantial increase in performance and compute densities (with orders of magnitude better performance see e.g., NVIDIA A100 vs.~P100 in FP16; the latter with \unit[18.7]{Tflop/s} peak).
This could, however, come at a cost and a potential loss in generality, which may be counter-productive from a HPC perspective.

\subsection{Matrix Engines: Motivation from HPC Side}\label{ssec:motivation-hpc}
The consideration for inclusion of MEs into future systems can be motivated by improvements in performance and energy efficiency by such a transition. Energy-consumption is one of the two limiting factors~\cite{shalf_exascale_2011}
(the other is cost) when building supercomputers, and any architectural choices that reduce energy (without sacrificing performance) will directly contribute to better systems. Furthermore, we can measure the impact of past architectural design-choices on energy consumption; choices that are similar to those we are looking at today. 

Consider, for example, exercising the Intel Xeon E5-2650v4 processor, which supports both scalar and vector instructions (AVX + AVX2) with a GEMM operation.
The energy-efficiency (in \unitfrac{flop~}{~J}) of this GEMM operation differs between the scalar and vectorized version and yields an average~2.3x observed increase in energy-efficiency, in favor of the vectorized version.
To measure this, we use OpenBLAS~\cite{openblas}, compiled with and without AVX support. 
OpenBLAS GEMM calls (for both single and double precision) are invoked with square matrices of size $n=5000$, and we repeat the call 30 times; resulting in a total of $2\cdot n^3 =$~\unit[7.5]{Tflop}. 
The energy consumption is measured using Intel Performance Counter Monitor (PCM) and we list the results in Table~\ref{tbl:cpupower}. 
The similar results for both DGEMM and SGEMM suggests that this increase in efficiency is agnostic to the numerical precision for this particular experiment.

\begin{table}[tbp]
    \caption{Energy-eff. of Vector Extensions on a Intel Xeon CPU}
    \centering\scriptsize
    \begin{tabularx}{\columnwidth}{|c|c|r|R|}
    \hline \hC
    \tH{\small Precision} & \tH{\small Vector extension} & \tH{\small Walltime} & \tH{\small Energy-efficiency} \\\hline
    \multirow{2}{*}{DGEMM}
        & |         & \unit[34.22]{s}       & \unit[1.23]{Gflop/J} \\ \cline{2-4} 
        &\cC AVX2   &\cC \unit[12.49]{s}    &\cC \unit[2.92]{Gflop/J} \\
    \hline
    \multirow{2}{*}{SGEMM}
        & |         & \unit[16.79]{s}       & \unit[2.65]{Gflop/J} \\ \cline{2-4} 
        &\cC AVX2   &\cC \unit[6.36]{s}     &\cC \unit[5.92]{Gflop/J} \\ \hline 
    \end{tabularx}
    \label{tbl:cpupower}
    \vspace{-4ex}
\end{table}

\begin{figure}[tbp]
    \centering
    \includegraphics[width=.8\columnwidth]{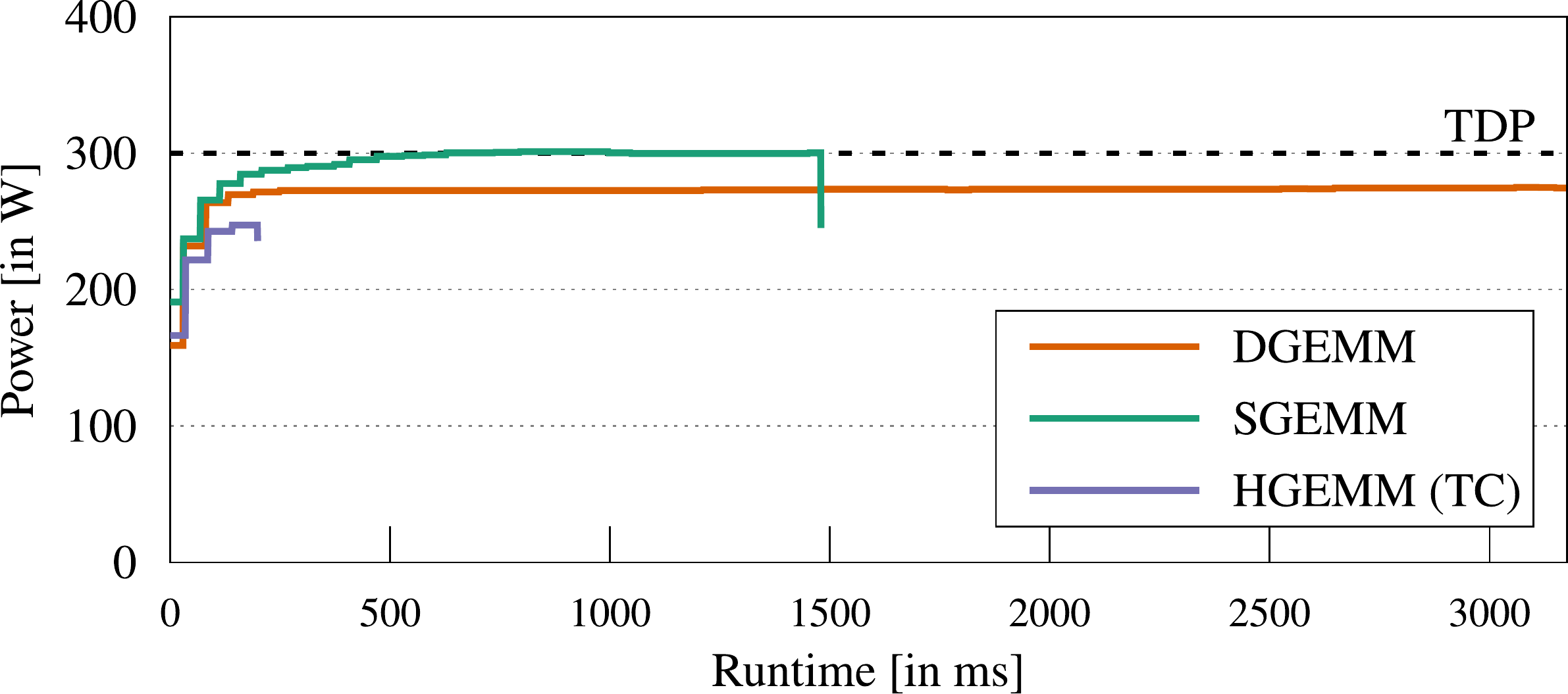}\vspace{1ex}
    \caption{Power consumption evaluation of GPU cores and TCs on a single Tesla V100 GPU using squared matrices of size $n=16384$; collected using NVML API calls (via nvmlDeviceGetPowerUsage).}
    \label{fig:power-gpu-tensorcore}
    \vspace{-2ex}
\end{figure}

Another example compares the energy efficiency of modern GPUs (in this case, an NVIDIA V100), which contains mixed-precision MEs. Executing a GEMM operation on such a processor, with and without using the built-in MEs, will lead to the observable performance and power consumption which we show in Figure~\ref{fig:power-gpu-tensorcore}. 
We observe that the accelerator's power can be greatly reduced using TCs. As Figure~\ref{fig:power-gpu-tensorcore} shows, we measure with square matrices of size $n=16384$ using TCs at half precision: see HGEMM (with TC), and GPU cores at single and double precision (see SGEMM \& DGEMM). Two important points to notea are: a) SGEMM and DGEMM draw power close to the TDP (\unit[300]{W}), and b) SGEMM or DGEMM cannot run concurrently with HGEMM. This indicates that SGEMM and DGEMM are already running at the highest possible performance (under TDP constraints), without any compromise due to resources occupied by the TCs.

Finally, with Moore's law ending, adding architecture support is no longer free, but comes at the expense of removing something else, and there is a large number of other architectural optimization that could yield better performance and/or energy efficiency for the same amount of silicon, e.g., improved caches, more aggressive OoO, more cores, etc. Hence, to reason whether spending silicon on MEs is a good architectural direction to pursue for HPC, we first must answer a more critical question: \textit{To what extent do HPC applications and DL workloads actually invoke GEMM-like operations?}

\section{Matrix Engines from a Software Perspective}\label{sec:software} 
In this section we describe, categorize, and analyze a broad set of
applications and benchmarks, from traditional HPC workloads
to modern Deep Learning, and analyze historical data from one supercomputer. The section is
structured as follows: first we perform a historical data and an offline
software dependency analysis, then conduct various measurements on AI and HPC
applications, and finally, for completeness, provide exact details about our
evaluation environments.

\subsection{Analysing Historical Data of the K Computer}\label{ssec:Klogs}
For the K Computer, RIKEN collected multiple
metrics of the executed application~\cite{yamamoto_k_2014} and the system,
such as power consumption and failure statistics. For parallel applications, launched via MPI, their database includes additional
information about the application binary's symbol table (through the Linux tool \texttt{nm}), but excludes data from linked shared libraries.

We analyze this data for a full year of K's operation (April~'18
to March~'19), during which the system executed $487,563$ scientific
applications distributed over 543 million node hours. Symbol table data is only
available for~96\% of these node hours, while the remainder can be attributed
to interactive jobs, non-parallel jobs, or jobs for which the collection was
intentionally disabled by the user. When searching the \texttt{nm} data for GEMM
functions, we can attribute~53.4\%, or $277,258,182$ node hours, to
applications which likely executed GEMM operations\footnote{Fujitsu's
compiler infrastructure defaults to selectively including individual functions
from their math kernel library, instead of linking the entire library.}.
However, the data does not reveal the exact node hours spent inside of GEMM routines. 

To conclude, in the absolute best case, the inclusion of MEs into the K computer could (in theory) have halved the number of node hours without any loss in quality for the applications, but a significant reduction in energy consumption (and, possibly, repair-costs), or an increase science throughput of the machine.

\subsection{Software Dependency Analysis via Spack}\label{ssec:spack}
The analysis of the job history of the K computer is rather narrow, both in
terms of common HPC workloads found in other countries as well as in terms of
possible math libraries providing and utilizing GEMM kernels.
Hence, we broaden the scope by also analyzing the software/library dependencies in Spack~\cite{gamblin_spack_2015}, which is a package manager that targets supercomputers. Spack provides easy access to thousands of scientific software packages to users without the necessity of the administrator's installation and maintenance.

We identified all libraries, which provide dense linear algebra routines, among
the~4,371 packages currently supported by Spack, namely: AMD BLIS, Atlas, BLIS,
Eigen, ESSL, Intel MKL, Netlib's LAPACK, ScaLAPACK and XBLAS, OpenBLAS, CUDA (cuBLAS), py-blis, libxsmm,
and veclibfort. Hereafter, we refer to these packages as BLAS libraries
with dependency distance~of~0. Using Spack's ability to track and list dependencies,
we subsequently identify packages which directly depend on libraries of
dependency distance~0, i.e., a set of packages with dependency distance of~1, and
so on and so forth. While the number of packages with dependency distance~0 is
only~14, the number increases to~239 in the next step (excluding the set of
dependency distance~0), or~5.47\% of all Spack packages.

Table~\ref{tab:spackblas}
lists further dependency distances, with the last column adjusting for the fact
that Spack includes a large number of sub-packages for Python and~R, which we
merge under their parent packages. As we can see, \textit{51\% (or 70\% without sub-package
adjustment) of Spack's packages depend directly or indirectly on BLAS libraries}.
While the similarity of the percentage with Section~\ref{ssec:Klogs} is mere
coincidental, our findings show that in the best case, only about half the packages could benefit from MEs. Unfortunately, our analysis is oblivious of a more important metric: \textit{How much time of the execution is actually spent in these BLAS routines?} In order to reason around said question, we split our efforts into two paths: \textbf{(i)}~What is the impact of using MEs in Deep Learning workloads, which are known to contain a lot of GEMM operations?, and \textbf{(ii)}~What \textit{could} be the impact of using MEs in regular HPC applications whose GEMM usage today remains unquantified?


\begin{table}[tbp]
    \caption{Dependency Analysis of Dense Linear Algebra Libraries for Spack
    Package Manager (w/ \& w/o Python and R sub-packages)}
    \centering\scriptsize
    \begin{tabularx}{\columnwidth}{|c|R|R|}
    \hline \hC
    \tH{Dependency Distance} & \tH{\# and \% of Packages} & \tH{excluding py-* \& R-*} \\ \hline
    $0$             & 14~~~(0.32)~    & 14~~~(0.55)~    \\ \hline \rC
    $1$             & 239~~~(5.47)~   & 226~~~(8.87)~   \\ \hline
    $2$             & 762~~(17.43)~   & 541~~(21.23)~   \\ \hline \rC
    $3$             & 968~~(22.15)~   & 714~~(28.02)~   \\ \hline\hline
    $1$--$\infty$   & 3061~~(70.03)~  & 1311~~(51.45)~  \\ \hline
    \end{tabularx}
    \label{tab:spackblas}
    \vspace{-4ex}
\end{table}

\subsection{Matrix Engines in Deep Learning Applications}
\label{ssec:ml}
Deep Learning (DL) has been a major driver in the active development of architectures with MEs, and corresponding software support, in recent years.
As a result, workloads of this class are ready for immediate empirical evaluation.
Moreover, since DL applications are typically comprised from a small number of well-known kernels (e.g., convolution), optimized implementations typically do not statically link to BLAS libraries.
They either decide at run-time to call BLAS depending on the kernel parameters, or inline matrix multiplications inside of kernels.
Therefore, we rely on empirical evaluation of DL applications, listed in the next Section, using our benchmarking framework\footnote{\textbf{Benchmarker} tool (available: \url{http://benchmarker.blackbird.pw/}) uses synthetic data sets, with same input sample size but reduced number of samples, to execute DL models on arbitrary number of GPUs; 1 GPU in our tests.} with PyTorch~\cite{pytorch} as the backend. 

\subsubsection{Machine Learning Benchmarks}\label{sssec:mlperf:apps}
For this evaluation we use a number of DL models spanning various domains and largely intersecting with the popular MLPerf benchmark~\cite{mlperf}.
Namely, we use:
BERT~\cite{devlin2019bert}, a language model based on self-attention mechanism; 
Cosmoflow~\cite{cosmoflow}, a 3D convolutional neural net (CNN) used in computational cosmology; 
VGG16~\cite{vgg} and ResNet50~\cite{resnet}, two 2D CNN for image recognition; 
DeepLabV3~\cite{deeplabv3}, a 2D CNN for image segmentation; 
SSD300~\cite{ssd300}, a 2D CNN for object detection;
and NCF~\cite{ncf}, a recommender model based on collaborative filtering.
Additionally, we benchmark several individual layers: GEMM/dense; LSTM~\cite{lstm} and GRU~\cite{gru}, both forms of recurrent units; 2D convolution layer; and self-attention~\cite{attention}. 


\subsubsection{Power Evaluation of DL Workloads}\label{sssec:mlperf:measurements}
\begin{figure}[tbp]
    \centering
    \includegraphics[width=.8\columnwidth]{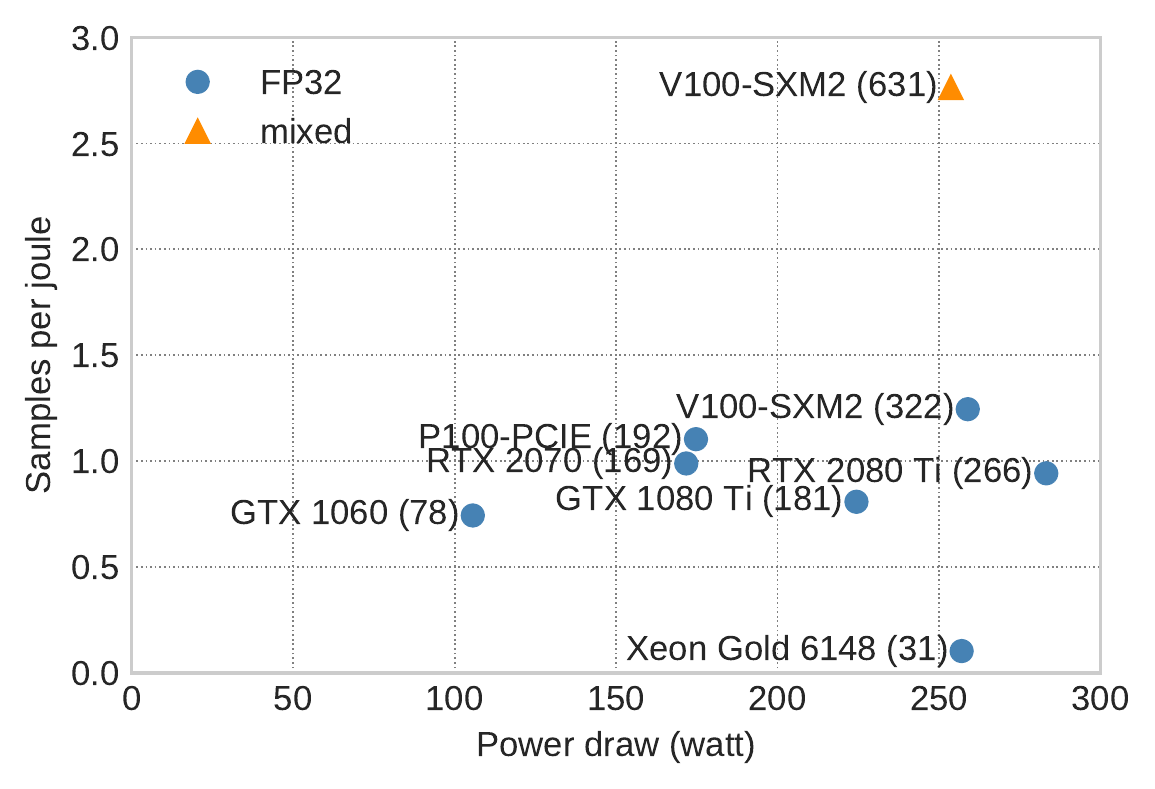}
    \caption{Energy-efficiency of ResNet50 training; Throughput in images/s (samples per second) indicated in parenthesis
    }
    \label{fig:dl-power-resnet}
    \vspace{-2ex}
\end{figure}
%
We first perform energy-efficiency measurements of ResNet50 training across a range of consumer and data-center chips (cf.~Section~\ref{ssec:hwforgpu} for further details).
Evaluation results are summarized in Figure~\ref{fig:dl-power-resnet}, which shows that while new GPUs are reported to be delivering better performance for DL applications (among other things), this increase comes at cost of higher power consumption while showing only a marginal generational improvement in energy-efficiency (\unitfrac{flop~}{~J}~; or image samples per Joule).
However, using TCs does yield higher energy efficiency (cf. \texttt{mixed} in Figure~\ref{fig:dl-power-resnet}), resulting from doubled image throughput at roughly the same power consumption.
This~2x practical improvement for ResNet50 is below our expected~$\approx$7x gain, derived from moving SGEMM operations to mixed precision HGEMM on TCs as visualized in Figure~\ref{fig:power-gpu-tensorcore}, and hence we analyze the fraction of GEMM operations in DL models in the next Section.

\subsubsection{Evaluation of Tensor Core Occupancy}\label{sssec:mlperf:gemm_measurements}
Table \ref{table:DL-TC-speedups} shows a range of DL models run with PyTorch framework in FP32 precision and in mixed precision mode, using apex library, on a Tesla V100 GPU.
We use NVIDIA's \texttt{nvprof} tool to collect function call profiles for individual kernels. 
It should be noted, that these results can be biased in both directions: for some kernels, e.g., 3D convolutions of Cosmoflow model, the implementation utilizing TCs is not yet available and the reported speedup is less than possible.
On the other hand, LSTM kernel running in reduced precision mode is utilizing a completely different underlying algorithm---thus the reported performance improvement can not be attributed to more efficient matrix multiplications.
However, after manually verifying which kernels are being executed, we consider these results to be representative of AI workloads in general.
To summarize, the investigated DL applications show performance improvements of 2x (typical ConvNets) to 4x (Transformers), when using TCs.
These numbers, while not as high as the 7.6x of pure GEMM kernels, still provide substantial runtime improvements to users and clearly validate the usage of TCs in the DL domain.
Furthermore, considering the percentage of execution time spend utilizing TCs, we expect even greater speedup with more efficient matrix engines.
\begin{table}[tbp]
    \caption{Throughput Improvement from FP32 to Mixed Precision. 
    \textbf{\%TC}: percentage of time spent on Tensor Cores (relative to total time);
    \textbf{\%TC comp}: compute time spent on TCs excluding data movement; and
    \textbf{\%Mem}: time for data movement between host and device.
    }
    \label{table:DL-TC-speedups}
    \centering\scriptsize
    \begin{tabularx}{\columnwidth}{|l|c|R|r|R|}
    \hline \hC
    \tH{\small Benchmark} & \tH{\small Speedup} & \tH{\small \% TC} & \tH{\small \% TC comp} & \tH{\small \% Mem} \\ \hline
    BERT      &      3.39x & 50.86 &      55.26 &   7.97 \\ \hline \rC
    Cosmoflow &      1.16x &  0.04 &       0.05 &  22.90 \\ \hline
    VGG16     &      1.71x & 12.30 &      12.74 &   3.45 \\ \hline \rC
    Resnet50  &      1.97x & 16.32 &      16.78 &   2.76 \\ \hline
    DeepLabV3 &      1.75x & 16.33 &      16.44 &   0.69 \\ \hline \rC
    SSD300    &      1.78x &  8.55 &       8.66 &   1.32 \\ \hline
    NCF       &      0.97x & 22.37 &      26.79 &  16.50 \\ \hline \rC
    GEMM      &      7.59x & 20.08 &      99.90 &  79.90 \\ \hline
    GRU       &      3.67x &  6.59 &       7.48 &  11.94 \\ \hline \rC
    LSTM      &      5.69x & 11.63 &      13.85 &  16.03 \\ \hline
    Conv2D    &      1.12x &  0.27 &       0.32 &  16.78 \\ \hline \rC
    Attention &      3.49x & 44.49 &      58.19 &  23.55 \\ \hline
    \end{tabularx}
    \vspace{-4ex}
\end{table}



\begin{table*}[tbp]
    \caption{\label{table:APPS} Overview of (Proxy-) Applications used for this Study; `(R)' for SPEC CPU indicates lack of OpenMP parallelization}
    \centering\scriptsize
    \begin{tabularx}{\textwidth}{l V{3} X|l V{3} X|l V{3} X|l|}
    \hline \hC
    \tH{\small Set} & \tH{\small Name} & \tH{\small Sci. / Eng. / AI Domain} & \tH{\small Name} & \tH{\small Sci. / Eng. / AI Domain} & \tH{\small Name} & \tH{\small Sci. / Eng. / AI Domain} \\ \hline
    \multirow{4}{*}{Deep Learning}
        & BERT & Natural Language Processing & 
        DeepLabV3 & Image Segmentation 
        & GRU & Single Layer\\ \cline{2-7}
        &\cC Cosmoflow  &\cC Computational Cosmology&
        \cC SSD300 &\cC Object Detection  &
        \cC LSTM &\cC Single Layer\\ \cline{2-7}
        & VGG16 & Image Recognition & 
        NCF & Recommender Systems& 
        Conv2D & Single Layer\\ \cline{2-7}
        &\cC Resnet50 &\cC Image Recognition
        &\cC GEMM &\cC Single Layer
        &\cC Attention &\cC Single Layer\\
    \midrule
    \multirow{1}{*}{TOP500}
        & HPL & Math/Computer Science & HPCG & Math/Computer Science & & \\
    \midrule
    \multirow{4}{*}{ECP}
        &\cC AMG &\cC Physics and Bioscience &\cC miniAMR &\cC Geoscience/Earthscience &\cC SW4lite &\cC Geoscience/Earthscience \\ \cline{2-7}
        & CoMD & Material Science/Engineering & miniFE & Physics & SWFFT & Physics \\ \cline{2-7}
        &\cC Laghos &\cC Physics &\cC miniTRI &\cC Math/Computer Science &\cC XSBench &\cC Physics \\ \cline{2-7}
        & MACSio & Math/Computer Science & Nekbone & Engineering (Mechanics, CFD) & & \\
    \midrule
    \multirow{3}{*}{RIKEN}
        &\cC FFB &\cC Engineering (Mechanics, CFD) &\cC mVMC &\cC Physics &\cC NTChem &\cC Chemistry \\ \cline{2-7}
        & FFVC & Engineering (Mechanics, CFD) & NGSA & Bioscience & QCD & Lattice QCD \\ \cline{2-7}
        &\cC MODYLAS &\cC Physics and Chemistry &\cC NICAM &\cC Geoscience/Earthscience &\cC &\cC \\
    \midrule
    \multirow{8}{*}{SPEC CPU}
        & blender(R) & Math/Computer Science & exchange2 & Artificial Intelligence & omnetpp & Math/Computer Science \\ \cline{2-7}
        &\cC cam4(R) &\cC Geoscience/Earthscience &\cC fotonik3d &\cC Physics &\cC perlbench &\cC Math/Computer Science \\ \cline{2-7}
        & namd(R) & Material Science/Engineering & gcc & Math/Computer Science & pop2 & Geoscience/Earthscience \\ \cline{2-7}
        &\cC parest(R) &\cC Bioscience &\cC imagick &\cC Math/Computer Science &\cC wrf &\cC Geoscience/Earthscience \\ \cline{2-7}
        & povray(R) & Math/Computer Science & lbm & Engineering (Mechanics, CFD) & roms & Geoscience/Earthscience \\ \cline{2-7}
        &\cC bwaves &\cC Physics &\cC leela &\cC Artificial Intelligence &\cC x264 &\cC Math/Computer Science \\ \cline{2-7}
        & cactuBSSN & Physics & mcf & Math/Computer Science & xalancbmk & Math/Computer Science \\ \cline{2-7}
        &\cC deepsjeng &\cC Artificial Intelligence &\cC nab &\cC Material Science/Engineering &\cC xz &\cC Math/Computer Science \\
    \midrule
    \multirow{5}{*}{SPEC OMP}
        & applu331 & Engineering (Mechanics, CFD) & fma3d & Physics & mgrid331 & Engineering (Mechanics, CFD) \\ \cline{2-7}
        &\cC botsalgn &\cC Bioscience &\cC ilbdc &\cC Engineering (Mechanics, CFD) &\cC nab &\cC Chemistry \\ \cline{2-7}
        & botsspar & Math/Computer Science & imagick & Math/Computer Science & smithwa & Bioscience \\ \cline{2-7}
        &\cC bt331 &\cC Engineering (Mechanics, CFD) &\cC kdtree &\cC Math/Computer Science &\cC swim &\cC Geoscience/Earthscience \\ \cline{2-7}
        & bwaves & Engineering (Mechanics, CFD) & md & Material Science/Engineering & & \\
    \midrule
    \multirow{5}{*}{SPEC MPI}
        &\cC $[$d$]$leslie3d &\cC Engineering (Mechanics, CFD) &\cC $[$l$]$GemsFDTD &\cC Physics &\cC socorro &\cC Material Science/Engineering \\ \cline{2-7}
        & $[$d$]$milc & Lattice QCD & lu & Engineering (Mechanics, CFD) & tachyon & Math/Computer Science \\ \cline{2-7}
        &\cC fds4 &\cC Engineering (Mechanics, CFD) &\cC $[$l$]$wrf2 &\cC Geoscience/Earthscience &\cC tera\_tf &\cC Geoscience/Earthscience \\ \cline{2-7}
        & GAPgeofem & Physics & pop2 & Geoscience/Earthscience & zeusmp2 & Engineering (Mechanics, CFD) \\ \cline{2-7}
        &\cC lammps &\cC Material Science/Engineering &\cC RAxML &\cC Bioscience &\cC &\cC \\ \hline
    \end{tabularx}
    \label{tab:appsoverview}
    \vspace{-3ex}
\end{table*}

\subsection{HPC Proxy-Applications and SPEC Benchmarks}\label{ssec:hpcapps}
In contrast to the previous section, here we are in looking into more traditional supercomputing and HPC applications.

\subsubsection{Application Overview and Input Selection}
We select benchmarks from six different sets which represent
typical workloads and are commonly used by HPC centers and vendors for architecture comparisons and hardware procurement:
\begin{itemize}[leftmargin=2.5mm]
    \item TOP500 Benchmarks~\cite{strohmaier_top500_2018}: The HPC community
    utilizes High Performance Linpack (HPL) and High Performance Conjugate Gradients (HPCG) for a world-wide performance ranking of supercomputers and HPC systems.
    \item Exascale Computing Project (ECP) Proxy Applications~\cite{exascale_computing_project_ecp_2018}: The ECP released with version 1.0 a set of~12 workloads, which are used for procurement by HPC centers in the USA. We investigate eleven of them in our study\footnote{We exclude CANDLE, since Section~\ref{ssec:ml} covers AI workloads extensively.}.
    \item RIKEN CCS' Fiber Miniapp Suite~\cite{riken_aics_fiber_2015}: The eight proxy applications developed by RIKEN CCS represent the priority areas of the Japanese government and were used in the procurement of Supercomputer Fugaku~\cite{matsuoka_a64fx_2019}.
    \item SPEC Benchmarks~\cite{standard_performance_evaluation_corporation_specs_2020}: While the SPEC corporation hosts benchmarks for numerous areas of interest, such as clouds or accelerators, we focus on SPEC CPU 2017~V1.1 (24 benchmarks; contains OpenMP-accelerated versions), and the two sets derived from HPC workloads, namely SPEC OMP 2012~V1.1 (14) and SPEC MPI 2007~V2.0.1 (18).
\end{itemize}
The detailed list of all 77 HPC benchmarks, i.e., individual names per set, as well as an overview of the scientific domain they are representing, is available
in Table~\ref{tab:appsoverview}.

Our selection of inputs and configurations for the individual benchmarks of the TOP500,
ECP, and RIKEN Fiber sets remains unchanged from our previous publication on
double-precision floating-point unit utilization, and therefore readers may
consult Sec.~II (B) of~\cite{domke_double-precision_2019} for further details.

SPEC benchmarks generally provide three different input sizes (ordered by short
to long runtime): \texttt{test}, \texttt{train}, and \texttt{ref}(erence). The
difference between those three options is usually the problem size, e.g., time
steps, grid size, \#atoms, etc., with a few exceptions as stated in the SPEC
documentation (yet we expect no major changes in compute patterns). Hence, we
select the \texttt{train} input configuration for all our SPEC measurements.
For SPEC MPI \texttt{mtrain} is preferred over \texttt{ltrain} whenever possible.
Furthermore, we setup SPEC to perform \texttt{peak} runs (not \texttt{base})
which allows the usage of OpenMP threads for SPEC CPU benchmarks.

\subsubsection{Measurement Methodology}\label{ssec:swmethod}
From our prior work~\cite{domke_double-precision_2019}, we know that TOP500,
ECP, and RIKEN's benchmarks are highly optimized proxies for the represented applications (with the sole exception of Laghos), which utilize high performance
math libraries, such as Intel's MKL whenever suitable. To identify the dense,
GEMM-like operation which could be accelerated by a ME, we employ the
profiling functionality of the Score-P performance analysis
tool~\cite{knupfer_score-p_2012}. We create a Score-P library wrapper for all
functions found in the header files for dense-matrix compute libraries of MKL, i.e.,
(C)BLAS, PBLAS, BLAS-like Extension Transposition Routines, (C)LAPACK, and
ScaLAPACK.

Furthermore, since the applications can exhibit considerable amounts
of initialization and post-processing phases, we identify these phases
and exclude them from profiling via Score-P API calls. Additionally, we search
the source codes for function names indicating GEMM operations or Fortran's
\texttt{matmul} intrinsic, and instrument them to be included in our application profiles for the TOP500, ECP and RIKEN's set.

Unfortunately, SPEC benchmarks are implemented without external dependencies
for portability and ease of use reasons. Hence, we are not able to rely
on our Score-P library wrapper for these three SPEC sets. Our workaround involves
a three step approach---besides searching the source codes for relevant
function names as above---before collecting the necessary metrics: first, we
utilize the Intel Advisor tool to perform a Roofline analysis and extract
all source code regions which are tagged as ``compute intensive'', meaning
functions and loops with sufficient Arithmetic Intensity, i.e., flop/byte
ratio~$\geq$~7 for our CPU testbed (cf.~System 1 in Section~\ref{ssec:hwforcpu}).
This list was further pruned by requiring a Point Weight
(PtW)~$\geq$~1, i.e., the self-elapse time of this region is at least~1\%
of the total elapse time of the benchmark. The next step consists of
\textit{two independent, manual code inspection of all~598 location}
to determine their compute pattern (majority vote; with third check if needed),
and instrument them to be included in our application
profiles when they perform a GEMM-like operation. We identify and
instrument~14 source code locations\footnote{Similarly, we identified numerous hand-written GEMM kernels in Nekbone.}.
And lastly, we rely on Score-P's automatic
compiler instrumentation and function filter ability---filtering the
same function names we collected for the MKL wrapper above---to profile
source code regions where the benchmark programmer replaced external
library calls with hand-written functions. For SPEC benchmarks, we omit
the exclusion of initialization and post-processing phases, since these
are negligible\footnote{According to SPEC's rules for benchmark submission:
95\% of runtime must be compute-bound and spend within the benchmark's
own source code.}.

For TOP500, ECP and RIKEN's benchmarks, we compile and execute the tests
equivalently (except for FFB\footnote{Using combination of Score-P and
Intel Compiler for FFB results in erroneous intermediate results and early
benchmark termination.}) to our previous work (i.e., using the Intel compiler suite,
compiler flags as listed in Sec.~III~(A), and a combination of MPI processes
and OpenMP threads as provided in Tab.~IV of~\cite{domke_double-precision_2019}).
These combinations yield the best performance for each application on our
benchmarking hardware.

In contrast, we primarily make use of the GNU compiler
suite for SPEC benchmarks\footnote{Only for GNU compilers, Score-P performs 
filtering of instrumented function at compile time, not at runtime,
reducing the profiling overhead.}, except for botsspar, bt331, fds4, perlbench,
and socorro, which require Intel's compilers to avoid runtime issues.
The main compiler optimizations are \texttt{-O3}, \texttt{-march=native} or
\texttt{-xHOST} (depending on the compiler suite), and additional flags for
vectorization, loop unrolling, and fast math\footnote{For compilation details
see our framework: \url{gitlab.com/domke/MEstudy}}. We run SPEC CPU and OMP
benchmarks consecutively and with~48 OpenMP threads to match the number of logical cores on our testbed. Similarly, all SPEC~MPI test are run
with~48 MPI processes matching the core count, because SPEC MPI benchmarks
are not OpenMP-parallelized. In the next section, we will provide the results
and analysis of our profiling measurements.

\subsubsection{Evaluation of Measurement Results}\label{ssec:hpceval} 
\begin{figure*}[tbp]
    \begin{center}
    \includegraphics[width=\textwidth]{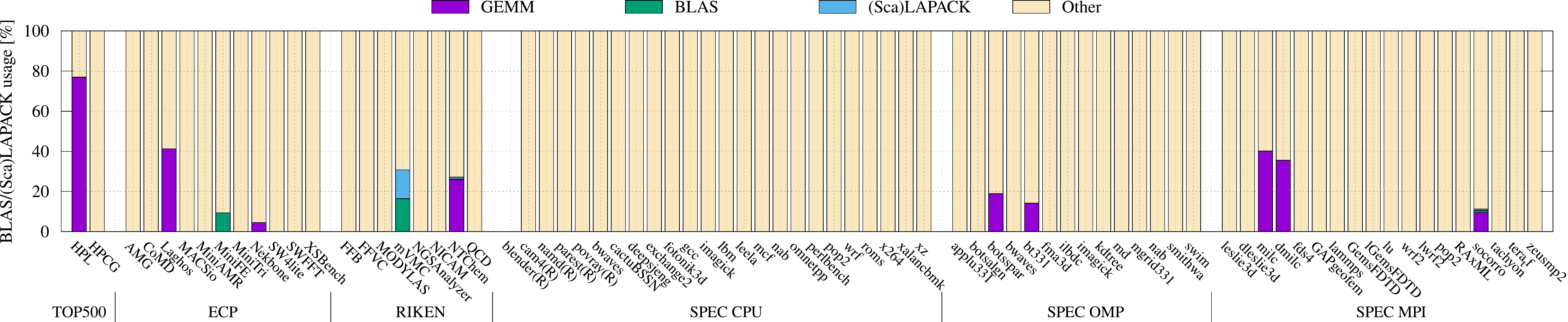}\vspace{1ex}
    \caption{GEMM, BLAS (non-GEMM functions), and (Sca)LAPACK utilization across 77 HPC benchmarks, see Section~\ref{ssec:hpceval} and  Table~\ref{tab:appsoverview} for details [Data for blender(R) missing due to unresolvable runtime errors; however, the source code contains GEMM function calls]}
    \label{fig:hpcgemm}
    \end{center}
    \vspace{-4ex}
\end{figure*}

The results of our measurements, as outlined in Section~\ref{ssec:swmethod},
to identify the GEMM usage per HPC (proxy-)application is visualized in
Figure~\ref{fig:hpcgemm}. We analyze the Score-P profiles for the runtime
spend in the profiled regions\footnote{Percentages
derived from ratio of \{region runtime\} to \{total application runtime minus
MPI\_Init/Finalize, and initialization/post-processing phases\}}.
Furthermore, we distinguish between, and show, four different compute regions
in the Figure: (1) GEMM operations, (2) BLAS operations (non-GEMM), i.e., all
BLAS Level[1$|$2$|$3] functions except for matrix-matrix multiplications,
(3) LAPACK and ScaLAPACK functions, and (4) all other code regions.
This distinction allows for an easy classification of code sections which can
be either directly accelerated~(1), or potentially indirectly accelerated~(2 \& 3),
or most probably not accelerated~(4) by MEs. To clarify, the
indirectly acceleration can result from use of GEMM operations within (Sca)LAPACK
routines or from mapping other BLAS function (such as GEMV) to MEs.

Unfortunately, as we can see from Figure~\ref{fig:hpcgemm}, the number of existing HPC
applications and workloads which can directly benefit from MEs is
rather sparse. Obviously, the most dominant example is High Performance Linpack.
In our test, HPL's compute kernel spends 76.81\% in GEMM operations and 0.14\%
in other BLAS function, and hence, executing the same test on a CPU with matrix
engine could substantially reduce the runtime of HPL. Other combinations of
benchmarks and chosen input parameters which perform GEMM are:
Laghos~(41.24\%), NTChem~(25.78\%), Nekbone~(4.58\%), SPEC OMP's botsspar~(18.9\%) and
bt331~(14.16\%), and SPEC MPI's milc \& dmilc~(40.16\% and 35.57\%, respectively) and
socorro benchmark~(9.52\%). We have instrumented GEMM regions in other
benchmarks as well, but our measurement indicate that
these regions are either dormant, or the type of input results in
alternative code paths, bypassing the GEMM operations.

Furthermore, we see that MiniFE, NTChem, mVMC, and socorro utilizes other BLAS 
functions during 9.38\%, 0.45\%, 16.41\%, and 0.99\% of their runtimes, respectively.
Upon closer inspection of the profiles, it is unlikely that the former two
benchmarks can benefit from MEs, because they utilize only BLAS Level 1 (i.e., 
vector-vector operations), while the other two also call BLAS Level 2 functions.
Looking at the (Sca)LAPACK dependencies and percentages of runtime in
Figure~\ref{fig:hpcgemm} yields similar results: mVMC spends noticeable amount
of runtime (14.35\%) in those libraries, while NTChem (0.95\%) and socorro's
(0.73\%) usage is negligible. Nevertheless, only these three out of all 77
could benefit from MEs if the used (Sca)LAPACK functions can be mapped to them.

Overall, we can state that only ten out of the 77 HPC benchmarks, which we
investigate at for this study, perform GEMM operation or outsource dense
linear algebra computations to libraries. Assuming an idealized equal
distribution of the node hours spend per our 77 benchmark on a supercomputer
results in an average runtime of 3.5\% spend in GEMM operations, or roughly
12 and a half days of a full year.

\subsection{Details of Evaluation Environments for Reproducibility}\label{ssec:hwswlist}
\subsubsection{CPU-based Measurements}\label{ssec:hwforcpu}
For the CPU-based measurement of our HPC (proxy-)applications in Section~\ref{ssec:hpcapps}, and the power efficiency evaluation in Section~\ref{ssec:motivation-hpc} we use a dual-socket x86\_64 compute node with Intel Xeon CPUs.
The details of this system are listed in Table~\ref{tab:hwconf} under \textbf{System 1}.

\subsubsection{GPU-based Measurements}\label{ssec:hwforgpu}
For the DL energy-efficiency measurements in Section~\ref{ssec:ml} we use NVIDIA GPUs ranging from consumer models, i.e., GTX~1060 \& 1080~Ti and RTX~2070 \& 2080~Ti, up to the data center lineup of Tesla P100-PCIE and V100-SXM2 GPUs. Additionally, the section also includes results of a Intel Xeon Gold CPU (compute node of ABCI~\cite{national_institute_of_advanced_industrial_science_and_technology_aist_ai_2020}; see \textbf{System 2} in Table~\ref{tab:hwconf}) for comparisons. 
The analysis of TC utilization by different kernels, see Table~\ref{table:DL-TC-speedups}, as well as power evaluation of GPU cores in Section~\ref{ssec:motivation-hpc}, and the comparisons between native and emulated precision GEMMs in Section~\ref{sec:emulating-high-precision}, is done on a Tesla V100-SXM2 of ABCI.


\subsubsection{Auxiliary Software}\label{ssec:auxsw}
We employ various auxiliary software packages to facilitate our measurements
on the different hardware architecture, such as compilers, Deep Neural Network
(DNN) libraries, and performance analysis frameworks. The main software
components are listed in Table~\ref{tab:swconf}.

\begin{table}[tbp]
    \caption{CPU-based Compute Nodes used for Measurements}
    \centering\scriptsize
    \begin{tabularx}{\columnwidth}{|X|l|l|}
    \hline \hC
    \tH{~}      & \tH{\small System 1~(\cref{ssec:motivation-hpc},~\cref{ssec:hpceval})}          & \tH{\small System 2~(\cref{sssec:mlperf:measurements})} \\ \hline
    Mainboard   & Supermicro X10DRG-Q           & Fujitsu Primergy-RX2540-M4 \\ \hline \rC
    CPU         & 2x Intel Xeon E5-2650v4       & Intel Xeon Gold 6148 \\ \hline
    \#Cores     & 24 (hyper-threading enabled)  & 20 (hyper-threading enabled) \\ \hline \rC
    Memory      & \unit[256]{GiB} DDR4 \unit[2400]{MHz} & \unit[32]{GiB} DDR4 \unit[2666]{MHz} \\ \hline
    OS          & CentOS Linux (v7.8.2003)      & CentOS Linux (v7.5.1804) \\ \hline \rC
    Kernel      & 3.10.0-1062                   & 3.10.0-862 \\ \hline
    \end{tabularx}
    \label{tab:hwconf}
    \vspace{-1ex}
\end{table}

\begin{table}[tbp]
    \caption{Auxiliary Software used for Measurements}
    \centering\scriptsize
    \begin{tabularx}{\columnwidth}{|r|l|r|l|}
    \hline \hC
    \tH{\small Package}    & \tH{\small Vers.}   & \tH{\small Package}  & \tH{\small Vers.} \\ \hline
    Intel Parallel Studio XE        & 2019; U1        & NVIDIA CUDA Toolkit           & 10.[1\&2]\\ \hline \rC
    Intel Advisor                   & 2020; U2\footnotemark & NVIDIA cuDNN & 7.6.5\\ \hline
    GNU Compiler Coll.         & 8.4.0                 & PyTorch ML Framework          & 7.6.5 \\ \hline \rC
    FUJITSU Software Suite          & 1.2.27b               & Score-P Analysis Frame.    & 6.0 \\ \hline
    Spack Package Manger            & 0.15.1                & Intel PCM tools   & 201710 \\ \hline
    \end{tabularx}
    \label{tab:swconf}
    \vspace{-3ex}
\end{table}
\footnotetext{Version 2019 (of Intel's P.S. XE) lacks needed feature for roofline analysis.}

\section{Gain from adapting Matrix Engines in HPC}
Assuming future CPUs and GPUs extensively integrate low (and/or high)
precision MEs, then we assess here the potential 
benefit for the HPC community when utilizing these units.

\subsection{Performance Extrapolation for Matrix Engines in HPC}
\label{sec:software:cost}

\begin{figure}[tbp]
    \centering
    \subfloat[K Computer]{
        \includegraphics[width=0.32\linewidth]{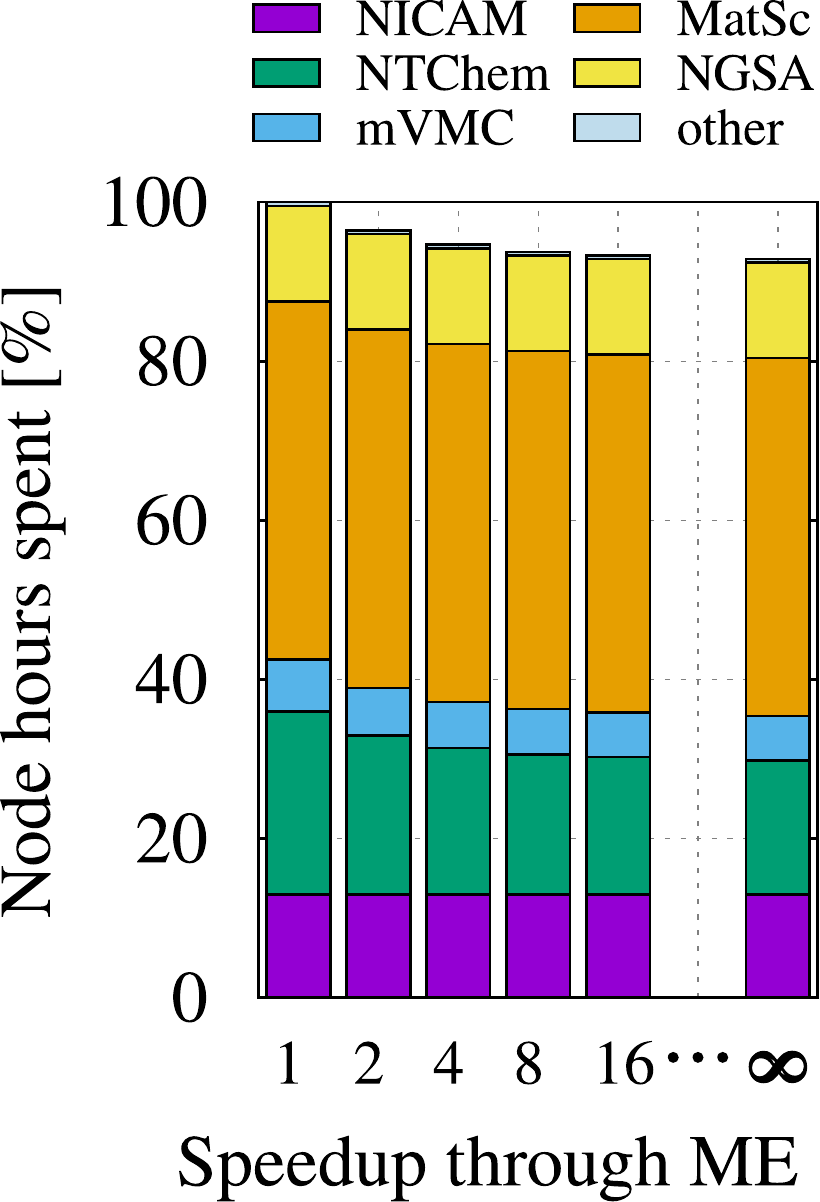}
        \label{fig:k_extrapol}
    }
    \subfloat[ANL systems]{
        \includegraphics[width=0.262\linewidth]{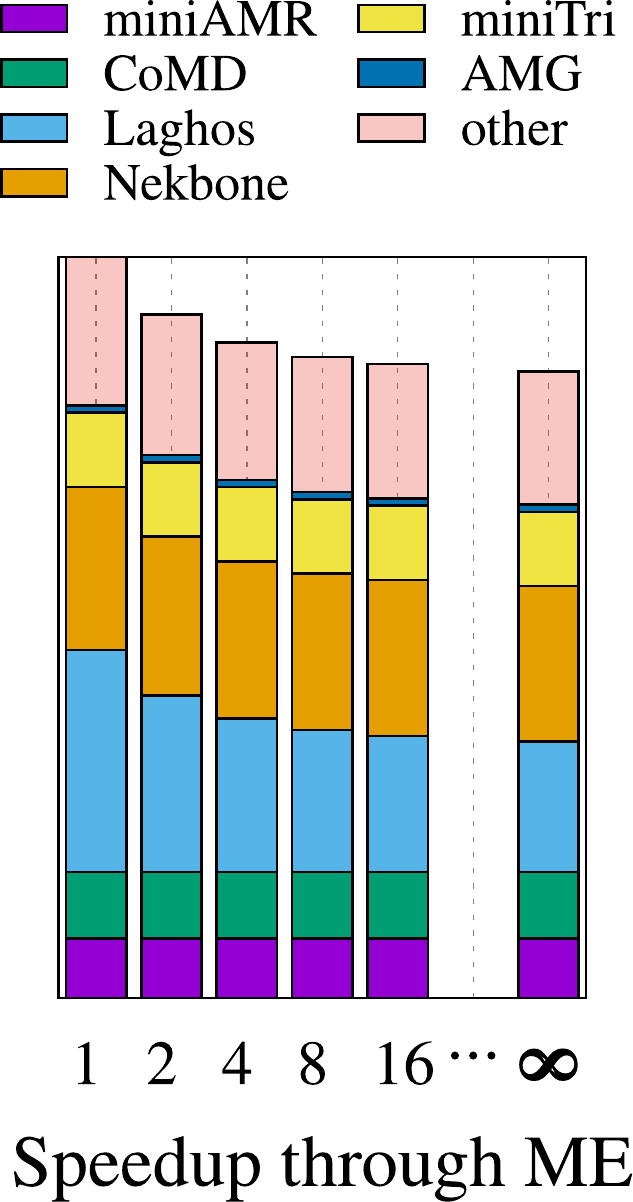}
        \label{fig:anl_extrapol}
    }
    \subfloat[Future System]{
        \includegraphics[width=0.314\linewidth]{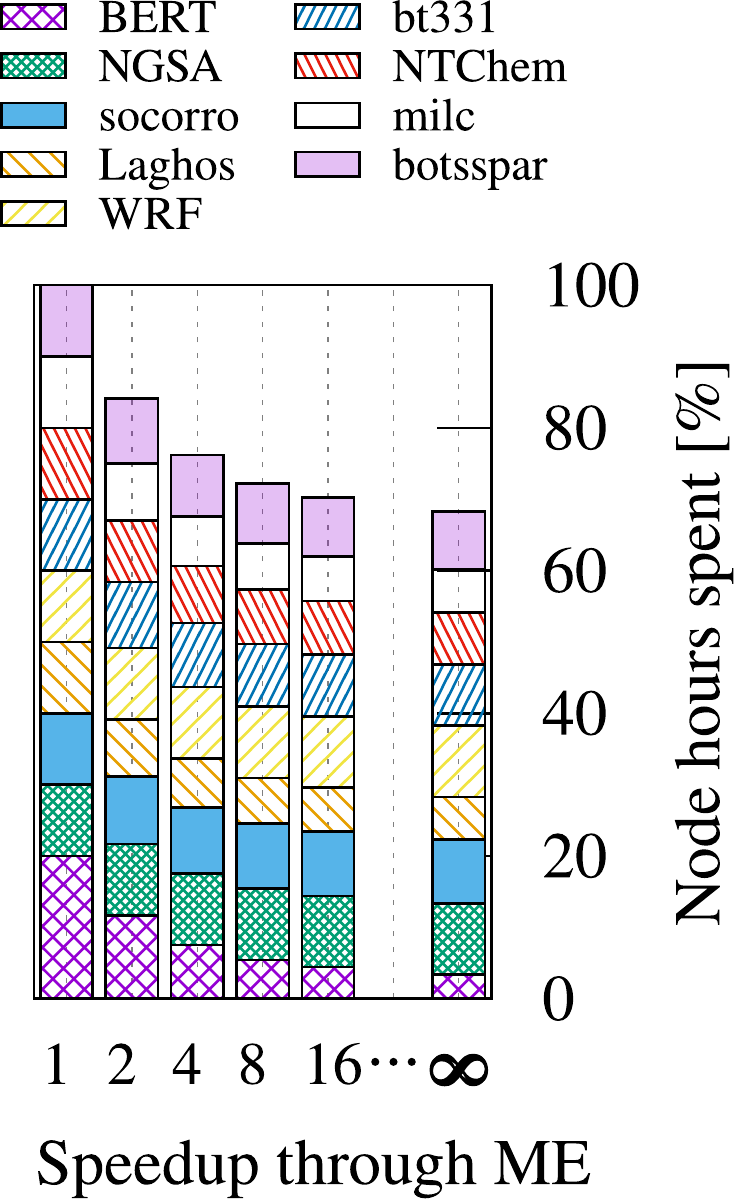}
        \label{fig:fiction_extrapol}
    }
    \caption{Node hour reduction by utilizing hypothetical ME; Breakdown of node hours per science domain based on historical data for Figure~\ref{fig:k_extrapol} and \ref{fig:anl_extrapol}; Future HPC system (\ref{fig:fiction_extrapol}) assumed to execute 20\% AI/DL tasks; In \ref{fig:k_extrapol}, \textbf{MatSc} represents FFB+MODYLAS+QCD in equal fractions}
    \label{fig:extrapol}
    \vspace{-2ex}
\end{figure}

From annual HPC system utilization reports for the K computer~\cite{aics_k_nodate}, and breakdown by domain, we know that the system was used primarily for material science (45\%), chemistry (23\%), geoscience (13\%), biology (12\%), and physics (6.5\%) calculations, plus 0.5\% ``other''. For the following thought experiment, we select for each science domain (see Table~\ref{tab:appsoverview}) a representative RIKEN Fiber benchmark, with FFB \& MODYLAS \& QCD representing material science equally (short MatSc), and we assume ``other'' applications spend 10\% in GEMM.
In Figure~\ref{fig:k_extrapol}, we visualize an extrapolation of node hours spend while assuming that these applications were accelerated by a ME for all GEMM and (Sca)LAPACK operations. Under these idealized conditions, and assuming a ME providing 4x speedup over baseline, the consumed node hours would have only reduced by~5.3\%. Even an infinitely fast ME would only yield a~7.1\% reduction.

Similarly, in Figure~\ref{fig:anl_extrapol}, we show the theoretically consumed node hours for the systems of the Argonne National Laboratory~(ANL)~\cite{collins_2016_nodate}, and we utilize our measurements from the ECP applications for the GEMM portion. Here, Laghos (representing 30\% physics simulations at ANL), Nekbone (rep.~22\% engineering), and ``other'' contribute to a 11.5\% reduction in node hours for a ME, which provides 4x speedup.

We perform the same extrapolation in Figure~\ref{fig:fiction_extrapol} for a fictional, future supercomputer. In this case, we assume that~20\% of its compute cycles are spend on AI/DL tasks, either within AI applications or as AI subroutine of another scientific application. The remaining 80\% are equally distributed across our eight science domains (see Table~\ref{tab:appsoverview}), and we select one representative application, with the highest GEMM and (Sca)/LACPACK percentage, for each.
For the AI portion, we select BERT and assume its GEMM occupancy is 83.2\%\footnote{The 83\% is derived from BERT's \texttt{\%TC comp} in Table~\ref{table:DL-TC-speedups} assuming the TCs yield 4x speedup over FP16 baseline using: $4 \cdot p / \big(4\cdot p + (100-p)\big) *100$}.
The inclusion of MEs, assuming they provide an 4x (or infinite) speedup, could reduce the node hours of our devised system by 23.8\% (or 32.8\%, respectively).

Obviously, these results in Figure~\ref{fig:extrapol} extrapolate the absolute best case scenario, since we sampled applications with the highest GEMM percentage, but in a real environment the node hour reduction is further constrained by non-GEMM applications (cf.~Figure~\ref{fig:hpcgemm}) and overheads, such as I/O or MPI.


\subsection{Emulating High-precision Compute with Low-precision MEs}\label{sec:emulating-high-precision}


While the industry is moving towards adopting MEs, whose impact on HPC we have analyzed in prior sections, there remains the issue regarding the numerical representation that these MEs will support.
While the recent NVIDIA A100 GPU does support full IEEE-754 double-precision in their MEs, other manufacturers may chose to provide MEs only for a shorter numerical representation (e.g., Intel Sapphire Roads supports bfloat16).
Are those low-precision MEs still valuable for use in future HPC systems? It turns out that they can be, as we can \textit{emulate} higher precision numerics using low-precision arithmetics.
While there exists several recent papers on the subject (e.g.,~\cite{markidis_nvidia_2018,8634417}), we will go into the details of one particular method called the Ozaki scheme~\cite{Ozaki:2012:ETM:2086820.2086827}.
It enables us to emulate a high-precision matrix-multiplication utilizing multiple low-precision matrix-multiplications.
For example, double-precision DGEMM can be computed using GEMM executed on Tensor Cores (cublasGemmEx)~\cite{10.1007/978-3-030-50743-5_12}.


In this Ozaki scheme, the mantissa and the exponent are computed separately. At a glance, firstly, the input matrices are element-wisely split into several matrices, starting with the one with the largest absolute value of the elements. The exponent part is separated at this point. Next, the all-to-all product of the split matrices is computed.
Here, the GEMM executed on TCs (cublasGemmEx) can be used for those matrix-multiplications. The exponent part is computed using integer operations. Finally, by summing the all-to-all product and restoring the exponent part, the accurate result of the matrix-multiplication is obtained.

Although this scheme uses FP64 arithmetic for the split and summation, cublasGemmEx dominates the total execution time. When the numbers of split matrices for matrices $A$ and $B$ are $s_A$, $s_B$, respectively, the most accurate result is obtained by performing $s_A \cdot s_B$ matrix-multiplications, but if only a DGEMM-equivalent accuracy is desired, the numbers of split matrices and matrix-multiplications can be reduced (cf.~\cite{10.1007/978-3-030-50743-5_12} for details). However, we note that the performance is input-dependent, i.e., the number of split matrices required depends on the absolute value range of the elements of the input matrix, the number of significand-bits in the elements, and the number of dimensions in the inner product direction of the matrices.

\begin{table}[tbp]
    \centering\scriptsize
    \caption{Performance of cuBLAS routines ($m=n=k=8192$) and GEMM-TC (software emulation using TCs) on Tesla V100.}
    \label{tab:emulation_perf}
    \begin{tabularx}{\columnwidth}{|l|l|r|R|r|}
    \hline \hC
    \tH{\small Implementation} & \tH{\small Condition} & \tH{\small \unit{Tflop/s}} & \tH{\small Watt} & \tH{\small Gflop/J} \\ \hline
    cublasGemmEx    & FP16/FP32-mixed           & 92.28 & 270.9 & 340.70 \\ \hline \rC
    cublasSgemm     & \multicolumn{1}{c|}{|}   & 14.54 & 276.1 &  52.66 \\ \hline
    cublasDgemm     & \multicolumn{1}{c|}{|}   &  7.20 & 286.5 &  25.14 \\ \hline\hline
    \multirow{3}{*}{SGEMM-TC}
        & input range: 1e+8     & 4.721     & 284.1     & 16.62 \\ \cline{2-5}
        &\cC input range: 1e+16 &\cC 2.140  &\cC 278.9  &\cC 7.67 \\ \cline{2-5}
        & input range: 1e+32    & 1.758     & 264.5     & 6.65 \\ \hline\hline
    \multirow{3}{*}{DGEMM-TC}
        & input range: 1e+8     & 1.097     & 273.7     & 4.01 \\ \cline{2-5}
        &\cC input range: 1e+16 &\cC 0.716  &\cC 271.4  &\cC 2.64 \\ \cline{2-5}
        & input range: 1e+32    & 0.618     & 270.4     & 2.29 \\ \hline
    \end{tabularx}
    \vspace{-3ex}
\end{table}

Table \ref{tab:emulation_perf} shows the computational performance of cuBLAS routines compared to emulated GEMM using TCs (cf.~[S$|$D]GEMM-TC) on a Tesla V100 ($m=n=k=8192$).
We approximate the \unit{flop/s} with $2\cdot n^3/\text{runtime}$, and measure the power consumption through NVIDIA's management library (NVML).
Although the performance of \mbox{GEMM-TC} does not outperform cuBLAS on this particular GPU, the techniques makes it possible to use TCs (or other matrix engines) to mitigate the lack of FP[32$|$64] MEs or to mitigate a limited availability of high-precision FPUs in other processors.
In fact, DGEMM-TC outperforms cublasDgemm on a NVIDIA Titan RTX, where 64-bit FPUs are limited~\cite{10.1007/978-3-030-50743-5_12}.

Other notable features of this scheme are: (1) bit-wise reproducibility (independent of the thread count and regardless of the rounding-error caused by GEMMs used in the computation), and (2) it can be used to compute dot-product and matrix-vector multiplication~\cite{ozblasppam2019}.
In the latter case, matrix engines could be used for the internal computations of the BLAS calls.



\section{Implications}\label{sec:implications} 
This section compiles a list of arguments for and against adding MEs to the compute systems designed for HPC.

\subsection{Opportunities for Replacing/Complimenting Vector Units}

\subsubsection{Dark Silicon}
As our experiments in Section~\ref{ssec:motivation-hpc} with TCs have shown, either of FPUs or TCs will get close to the TDP of the device; and both FPUs and TCs can not be used at the same time.
Accordingly, one could treat the chip area taken by TCs as simply non-valuable; even if one would release this chip area used by TCs, we are not able to add other resources to contribute to compute performance coming from the FPUs without changing the TDP (since both SGEMM and DGEMM are already running close to TDP as Figure~\ref{fig:power-gpu-tensorcore} shows). 


\subsubsection{Other Compute Patterns Benefiting from Matrix Engines}
The use of MEs can be expanded to compute patterns that could be represented as dense matrix operations. Note that the use of MEs in Deep Learning is driven by re-structuring convolution filters into matrices~\cite{lecun2015deep}, an approach which may soon be obsolete~\cite{hoefler_sparsity_2021}. In HPC, similar examples are emerging, e.g., efforts include accelerating sparse matrix-matrix multiplication by fitting tiles of the sparse matrix which contain one or more elements to TC fragments~\cite{Zachariadis_2020}. Other efforts try to exploit MEs at a lower level by using compiler-based approaches (i.e. polyhedral analysis) to automatically transform compute-intense nest loops to TCs~\cite{bhaskaracharya2020automatic}.

\subsubsection{Lower/Mixed Precision in Scientific Computing}
MEs can be designed to have hardware support for arbitrary precisions (as discussed in Section~\ref{sec:hardware}). However, the demand from AI/ML driving the rise of MEs incentivizes vendors to use lower precision. Lower precisions that are gaining more support in MEs could in turn incentivize the HPC community to re-evaluate its stance on precision requirements (we refer the reader to a detailed survey that elaborates on the opportunities of mixed precision in commonly used numerical methods~\cite{abdelfattah2020survey}).

\subsection{Challenges for Replacing Vector Units with Matrix Engines}
\subsubsection{Inefficiency for Level-1 and Level-2 BLAS Operations}
Current MEs are mostly built using systolic arrays (cf.~Section~\ref{sec:hardware}). The 2D nature of systolic arrays makes them extremely efficient for matrix-matrix multiplication (Level-3 BLAS operations), but not very-efficient at other operations such as Matrix-Vector (Level-2 BLAS), or Vector-Vector (Level-1 BLAS) given one of the dimensions of the systolic array would be waiting for the vector to propagate through. SIMD-style designs meanwhile can work efficiently for BLAS Levels 1 \& 2~\cite{4772956}.
\subsubsection{Programmability Burden}
The use of MEs in AI/ML is to a large extent hidden from the end users, i.e.\ happens inside the frameworks and libraries. The use of MEs for dense matrix operators is also hidden behind APIs provided by linear algebra libraries (e.g., NVIDIA exposes a  API that enables linear algebra libraries to use TCs). However, for other compute patterns, unsupported by such libraries, it becomes the burden of the user to write source code to use the MEs. Approaches for auto-generation of code for MEs~\cite{bhaskaracharya2020automatic} are still in very early stages, and will take time to mature. 
\subsubsection{Reduced Portability}
Portability of source code between SIMD vector extensions from different generations is often troublesome and requires manual tuning of the code (e.g. using \texttt{\#ifdef} programming style). We have no historical evidence indicating that the situation will be any different with MEs---if not more complicated, given that there no convergence yet on how MEs are exposed to the end user and compiler.
\subsubsection{Is the Dark Silicon Effect General?}
As we showed, TCs in NVIDIA GPUs do not take away from the performance potential of 32-$|$64-bit FPUs, which run close to TDP. However, there is no clear evidence that this dark silicon effect would be the same for CPUs, or GPUs made by other vendors.

\subsubsection{Overhead of Data Staging to Matrix Engine}
When the compiler vectorizes a loop, vector registers are used for the operands. Accordingly, vectors extensions do not have any particular overhead for data movement (or staging). In comparison, TCs/MEs, in the currently available forms, come with an overhead attributed to the off-loading model that separates the ME memory hierarchy from the host processor (CPU or GPU).

\section{Related Work}\label{sec:relatedwork} 

Ahmad et al.~\cite{abdelfattah2020survey} conducted a comprehensive survey---from the point of view of numerical methods---of efforts that utilize mix-precision hardware.
The survey focuses on dense linear algebra methods such as HGEMM and LU factorization, communication compression using reduced precision, sparse algorithms, and other algorithms that are fundamental parts of HPC applications.
They also provided an overview of existing math libraries that support mixed-precision arithmetic.

Markidis et al.~\cite{markidis_nvidia_2018} proposed a method to achieve the SGEMM-equivalent accuracy TC-accelerated GEMMs. Additionally, the authors provided a method to cope with precision loss. 
Zachariadis et al.~\cite{Zachariadis_2020} focus on mixed-precision sparse general matrix-matrix multiplication (spGEMM). 
Furthermore, Mukunoki et al.~\cite{10.1007/978-3-030-50743-5_12} proposed a method to utilize TCs to emulate SGEMM and DGEMM, and Sorna et al.~\cite{8634417} proposed a method to improve the accuracy of 2D fast Fourier transformation on TC. Overall, those efforts were directed to specific use cases of Tensor Cores, and do not have the HPC-wide perspective of MEs which we provide in this study.

\section{Conclusion}\label{sec:conclusion} 

To conclude, we revisit our introductory questions.
Our detailed analysis of the prospect of matrix engines, both for Deep Learning and for HPC workloads, gives little reason to believe that the HPC community at large should actively embrace and pursue the inclusion of more, faster, and higher-precision matrix engines.
For DL the current architectures, such as Tensor Cores, will likely soon hit a point of diminishing returns due to Amdahl's Law and the research will likely be tilting towards more sparsity, while for traditional HPC the low utilization of GEMM does not seem to merit a wide adoption.

Furthermore, we explore opportunities that could arise from the availability of matrix engines in future systems, which signals that the supercomputing community should be careful not to prematurely dismiss MEs, specially as they proliferate due to market forces.
Obviously, individual HPC centers need to revisit their particular priority applications to make a final assessment, and potentially an overall science throughput improvement of $\approx$1.1x, as we demonstrate for existing supercomputers, might justify the investment if all other architectural options have been exhausted.






\section*{Acknowledgment}

This work was supported by
the Japan Society for the Promotion of Science KAKENHI Grant Number 19K20286;
by JST, PRESTO Grant Number JPMJPR20MA, Japan;
by the New Energy and Industrial Technology Development Organization (NEDO);
and the AIST/TokyoTech Real-world Big-Data Computation Open Innovation Laboratory (RWBC-OIL).
This research utilized the Cygnus HPC system, provided by the Multidisciplinary Cooperative Research Program of the Center for Computational Sciences, University of Tsukuba.

\bibliographystyle{IEEEtran}
\bibliography{IEEEabrv,main}

\end{document}